 	\def\version{May 4, 2004}

\newif\ifpdf
\ifx\pdfoutput\undefined
\pdffalse 
\else
\pdfoutput=1 
\pdftrue \fi

\newif\iffinal
\finalfalse	
\finaltrue 

\documentclass[reqno,twoside,11pt]{amsart}
\iffinal\else\usepackage[notref,notcite]{showkeys}\fi
\usepackage{cite}
\usepackage{amsmath}
\usepackage{amsfonts}
\usepackage{amssymb}
\usepackage{graphicx}
\usepackage{graphics}
\usepackage{verbatim}

\IfFileExists{epsf.def}{\input epsf.def}{\usepackage{epsf}}

\IfFileExists{myowntimes.sty}{\usepackage{myowntimes}\usepackage{mathrsfs}}
	{\usepackage{times}\newcommand{\mathscr}{\mathcal}}
\DeclareFontFamily{OT1}{eusb}{} \DeclareFontShape{OT1}{eusb}{m}{n}
{<5> <6> <7> <8> <9> <10> <11> <12> <14.4> eusb10}{}
\DeclareMathAlphabet{\eusb}{OT1}{eusb}{m}{n}

\DeclareFontFamily{OT1}{eusm}{} \DeclareFontShape{OT1}{eusm}{m}{n}
{<5> <6> <7> <8> <9> <10> <11> <12> <14.4> eusm10}{}
\DeclareMathAlphabet{\eusm}{OT1}{eusm}{m}{n}

\DeclareFontFamily{OT1}{eufm}{} \DeclareFontShape{OT1}{eufm}{m}{n}
{<5> <6> <7> <8> <9> <10> <11> <12> <14.4> eufm10}{}
\DeclareMathAlphabet{\mathfrak}{OT1}{eufm}{m}{n}

\DeclareFontFamily{OT1}{fraktura}{}
\DeclareFontShape{OT1}{fraktura}{m}{n} {<5> <6> <7> <8> <9> <10>
<11> <12> <13> <14.4> [1.1] eufm10}{}
\DeclareMathAlphabet{\fraktura}{OT1}{fraktura}{m}{n}

\DeclareFontFamily{OT1}{cmfi}{} \DeclareFontShape{OT1}{cmfi}{m}{n}
{<5> <6> <7> <8> <9> <10> <11> <12> <13> <14.4> [0.9] cmfi10}{}
\DeclareMathAlphabet{\cmfi}{OT1}{cmfi}{b}{n}

\DeclareFontFamily{OT1}{cmss}{} \DeclareFontShape{OT1}{cmss}{m}{n}
{<5> <6> <7> <8> <9> <10> <11> <12> <13> <14.4> cmss10}{}
\DeclareMathAlphabet{\cmss}{OT1}{cmss}{m}{n}

\newtheoremstyle{thm}{1.5ex}{1.5ex}{\itshape\rmfamily}{}
{\bfseries\rmfamily}{}{2ex}{}

\newtheoremstyle{def}{1.5ex}{1.5ex}{\rmfamily}{}
{\bfseries\rmfamily}{}{2ex}{}

\newtheoremstyle{rem}{1.3ex}{1.3ex}{\rmfamily}{}
{\itshape}
{} {1.5ex}{}

\newenvironment{proofsect}[1]
{\vskip0.1cm\noindent{\rmfamily\itshape#1.}}{\qed\vspace{0.15cm}}

\theoremstyle{thm}
\newtheorem{theorem}{Theorem}[section]
\newtheorem{lemma}[theorem]{Lemma}

\newtheorem*{Main Theorem}{Main Theorem.}

\theoremstyle{def}
\newtheorem{definition}{Definition}

\theoremstyle{rem}
\newtheorem{remark}{{\itshape Remark}}[]

\numberwithin{equation}{section}

\renewcommand{\section}{\secdef\sct\sect}
\newcommand{\sct}[2][default]{\refstepcounter{section}
\addcontentsline{toc}{section}
{{\tocsection {}{\thesection}{\!\!\!\!#1\dotfill}}{}}
\vspace{0.7cm}
\centerline{ 
\scshape\arabic{section}.\ #1} \nopagebreak \vspace{0.2cm}}
\newcommand{\sect}[1]{
\vspace{0.4cm} \centerline{\large\scshape\rmfamily #1}
\vspace{0.2cm}}

\renewcommand{\subsection}{\secdef\subsct\sbsect}
\newcommand{\subsct}[2][default]{\refstepcounter{subsection}
\addcontentsline{toc}{subsection}
{{\tocsection{\!\!}{\hspace{1.2em}\thesubsection}{\!\!\!\!#1\dotfill}}{}}
\nopagebreak\vspace{0.45\baselineskip} {\flushleft\bf
\arabic{section}.\arabic{subsection}~\bf #1.~}
\\*[3mm]\noindent
\nopagebreak}
\newcommand{\sbsect}[1]{\vspace{0.1cm}\noindent
\textbf{#1.~}\vspace{0.1cm}}

\renewcommand{\subsubsection}{%
\secdef \subsubsect\sbsbsect}
\newcommand{\subsubsect}[2][default]{%
\refstepcounter{subsubsection} 
\addcontentsline{toc}{subsubsection}{{\tocsection{\!\!}
{\hspace{3.05em}\thesubsubsection}{\!\!\!\!#1\dotfill}}{}}
\nopagebreak
\vspace{0.15\baselineskip} \nopagebreak {\flushleft\rmfamily
\itshape\arabic{section}.\arabic{subsection}.\arabic{subsubsection}
\ \rmfamily #1\/.}\ }
\newcommand{\sbsbsect}[1]{\vspace{0.1cm}\noindent
\rmfamily \itshape
\arabic{section}.\arabic{subsection}.\arabic{subsubsection} \
\sffamily #1\/.\ }
\iffinal
\newcommand{\printversion}{}
\else
\newcommand{\printversion}{, \version}
\fi

\newcommand{\textd}{\text{\rm d}\mkern0.5mu}
\newcommand{\texti}{\text{\rm i}\mkern0.7mu}

\renewcommand{\AA}{\mathcal A}
\newcommand{\BB}{\mathcal B}

\newcommand{\GG}{\mathcal G}

\newcommand{\E}{\mathbb E}

\newcommand{\BbbP}{\mathbb P}

\newcommand{\R}{\mathbb R}

\newcommand{\T}{\mathbb T}

\newcommand{\Z}{\mathbb Z}

\newcommand{\twoeqref}[2]{(\ref{#1}--\ref{#2})}
\newcommand{\1}{{1\mkern-4.5mu\textrm{l}}}
\renewcommand{\1}{\text{\sf 1}}

\newcommand{\scrK}{\mathscr{K}}

\newcommand{\bS}{\boldsymbol S}
\newcommand{\bt}{\boldsymbol t}

\newcommand{\btheta}{\boldsymbol\theta}
\newcommand{\bvartheta}{\boldsymbol\vartheta}

\newcommand{\bzero}{\boldsymbol 0}

\newcommand{\Jt}{J_{\text{\rm t}}}

\newcommand{\bk}{\text{\bfseries\itshape k}\mkern1mu}
\newcommand{\br}{\text{\bfseries\itshape r}\mkern1mu}

\newcommand{\hate}{\hat{\text{\rm e}}}

\newcommand{\cc}{{\text{\rm c}}}

\newcommand{\BBE}{\BB_{\text{\rm E}}}
\newcommand{\BBSW}{\BB_{\text{\rm SW}}}

\begin{document}

\title[Order by disorder in an $O(2)$-spin system\printversion]
{\Large Order by disorder, without order,\\in a two-dimensional spin system\\with \textit{O(2)} symmetry}

\author[M.~Biskup, L.~Chayes and S.A.~Kivelson\printversion]
{Marek~Biskup,${}^1$\, Lincoln~Chayes${}^1$ \,and\, Steven A.~Kivelson${}^2$}

\thanks{\hglue-4.5mm\fontsize{9.6}{9.6}\selectfont\copyright\,2004 by M.~Biskup, L.~Chayes, S.A.~Kivelson. Reproduction, by any means, of the entire article for non-commercial purposes is permitted without charge.\vspace{2mm}}
\maketitle

\vspace{-5mm}
\centerline{${}^1$\textit{Department of Mathematics, UCLA, Los Angeles, California, USA}}
\centerline{${}^2$\textit{Department of Physics, UCLA, Los Angeles, California, USA}}

\vspace{2mm}
\begin{quote}
\footnotesize \textbf{Abstract:}
We present a rigorous proof of an ordering transition for a two-component two-dimen\-sio\-nal antiferromagnet with nearest and next-nearest neighbor interactions. The low-temperature phase contains two states distinguished by local order among columns or, respectively, rows. Overall, there is no magnetic order in accord with the classic Mermin-Wagner theorem. The method of proof employs a rigorous version of ``order by disorder,'' whereby a high degeneracy among the ground states is lifted according to the differences in their associated spin-wave spectra. 
\end{quote}
\vspace{2mm}

\section{Introduction}
\label{sec1}\vspace{-3mm}
\subsection{Background}
\label{sec1.1}\noindent
For two-dimensional spin systems, the celebrated Mermin-Wagner theorem~\cite{Mermin-Wagner,Mermin} (and its extensions~\cite{DS,ISV}) precludes the possibility of the spontaneous breaking of a continuous internal symmetry.
However, this result does not prevent such models from exhibiting phase transitions. For example, in the usual XY-model there is a low-temperature phase, known as the~Kosterlitz-Thouless phase~\cite{Kosterlitz-Thouless}, characterized by power-law decay of correlations and, of course, vanishing spontaneous magnetization~\cite{McBryan-Spencer,Frohlich-Spencer}. The existence and properties of this phase have been of seminal importance for the understanding of various low-dimensional physical phenomena, e.g.,~2D~superconductivity and superfluidity, 2D~Josephson arrays, 2D~melting, etc. It it widely believed that no such phase exists for~$O(n)$-models with~$n\ge3$ although rigorous arguments for (or against) this conjecture are lacking.

Of course, among such models there are other pathways to phase transitions aside from attempting to break the continuous symmetry.
One idea is to inject additional \emph{discrete} symmetries into the model and observe the breaking of these ``small'' symmetries regardless of the (global) status of the ``big'' one. 
As an example, at each~$\br\in\Z^d$ (where~$d\ge2$) let us place a pair $(\sigma_{\br},\pi_{\br})$ of $n$-component unit-length spins whose interaction is described by the Hamiltonian
\begin{equation}
\mathscr{H}=-J_1\sum_{\langle\br,\br'\rangle}
(\sigma_{\br}\cdot\sigma_{\br'}+\pi_{\br}\cdot\pi_{\br'})
-J_2\sum_{\br}(\sigma_{\br}\cdot\pi_{\br})^2,
\end{equation}
where~$\langle\br,\br'\rangle$ denotes a pair of nearest neighbors on~$\Z^d$ and~$J_1,J_2>0$.
Obviously, this model has~$O(n)$ symmetry (rotating all spins) as well as a discrete~$\Z_2$ symmetry (relative reflection between the~$\sigma$'s and the~$\pi$'s). It is not hard to show that at low temperatures, regardless of the global status of the~$\sigma$'s and~$\pi$'s, there is coexistence between a phase where the~$\sigma$'s and~$\pi$'s are locally aligned with one another and one where they are locally antialigned. (Note that this is based purely on energy considerations---the said alignments are the only minimizers of the second term in the Hamiltonian.) A model similar to the one defined above was analyzed in~\cite{Senya1} where the corresponding conclusions were indeed established. We remark that these results hold even if~$d=2$ (and even if~$n>2$).

Another ``circumvention'' is based on the adaptation of the large-entropy methods to systems which happen to have continuous symmetry. These are distinguished from the more commonly studied systems in and of the fact that there is no apparent \emph{order parameter} signaling the existence of a low-temperature phase. The key idea dates back to~\cite{Kotecky-Shlosman,DS2} where some general principles were spelt out that guarantee a \emph{point} of phase coexistence.
Let us consider an attractive system where there is an energetically favored alignment which confines the spin configurations to a small portion of the spin space. Suppose that there are many other less favored alignments with an approximately homogeneous energy. Under these conditions, a first-order transition at some (intermediate) value of temperature is anticipated. This kind of transition was established for specific systems (including the~$q$-state Potts model) in~\cite{Kotecky-Shlosman,DS2}, see also~\cite{Senya2}. The general philosophy can easily be adapted to spin systems with a continuous symmetry, e.g., as in~\cite{Alexander-Chayes,CKS,CSZ} where some related problems were~discussed.

To illustrate these matters let us consider an example from~\cite{Alexander-Chayes}. Here we have a two-component spin of length one at each site of~$\Z^2$ which we parametrize by an angular variable~$\theta_{\br}\in(-\pi,\pi]$. Let~$V(x)$ denote the function which equals negative one if~$|x|<\epsilon$ and zero otherwise, and let
\begin{equation}
\mathscr{H}=J\sum_{\langle\br,\br'\rangle}V(\theta_{\br}-\theta_{\br'}),
\end{equation}
where, of course, the arguments of~$V$ are interpreted modulo~$2\pi$.
Then, at some parameter value~$J=\Jt$ obeying~$e^{\Jt}\approx\sqrt\epsilon$, coexistence occurs between a phase where nearly all neighboring spins are closely aligned and one where, locally, spins exhibit hardly any correlation. We reiterate that the use of~$n=2$ and $d=2$ is not of crucial importance for proofs of statements along these lines. Indeed, in~\cite{vanEnter-Shlosman1,vanEnter-Shlosman2}, similar results have been established in much generality. 

In all of the above examples a moment's thought reveals that no violation of the Mermin-Wagner theorem occurs. Indeed, this theorem does not preclude a phase transition, it only precludes a phase transition which is characterized by breaking of a (compact) continuous internal symmetry.

\subsection{Foreground}
\label{sec1.2}\noindent
The purpose of this note is to underscore another route ``around'' the Mermin-Wagner theorem.
The distinction here, compared to all of the abovementioned, is that it may take the reader \emph{two} moments to realize that our results are also in accord with the Mermin-Wagner theorem. Not unrelated is the fact that in our example the mechanism for ordering is relatively intricate.
Let us go right to the (formal) Hamiltonian which reads
\begin{equation}
\label{Ham}
\mathscr{H}=J\sum_{\br}\bigl(\bS_{\br}\cdot\bS_{\br+\hate_x+\hate_y}
+\bS_{\br}\cdot\bS_{\br+\hate_x-\hate_y}\bigr)
+J\gamma\sum_{\br}\bigl(\bS_{\br}\cdot\bS_{\br+\hate_x}
+\bS_{\br}\cdot\bS_{\br+\hate_y}\bigr).
\end{equation}
Here~$\br$ denotes a site in~$\Z^2$ and the~$\bS_{\br}$ are unit-length two-component spins, i.e.,~$\bS_{\br}\in\R^2$, with $|\bS_{\br}|=1$, for each~$\br\in\Z^2$. The vectors~$\hate_x$ and~$\hate_y$ are unit vectors in the~$x$ and~$y$ lattice directions while~$J$ (the overall interaction strength) and~$\gamma$ (the relative strength of nearest neighbor couplings) are positive numbers.
Notice the sign of the coupling---there is antiferromagnetism all~around.

In order to analyze the ground states, let us focus on the cases~$\gamma\ll1$. (Later we will only require~$\gamma<2$.) Notice, especially in this limit, that the interaction splits the lattice into even and odd sublattices. For the ground-state problem, say in an even-sided finite volume with periodic boundary conditions, it is clear that both of the sublattices will be Ne\'el (i.e., antiferromagnetically) ordered. However, once this Ne\'el order is in place, it is clear that the energetics are insensitive to the relative orientation of the spins on the two sublattices. Specifically, the spin at any site~$\br$ couples antiferromagnetically to the \emph{sum} of $\bS_{\br+\hate_x}$, $\bS_{\br+\hate_y}$, $\bS_{\br-\hate_x}$ and $\bS_{\br-\hate_y}$ which, in any Ne\'el state, is exactly zero. Thus we conclude that the set of ground states, i.e., the ``order-parameter space,'' cf~\cite{Mermin2}, of this model exhibits an $O(2)\otimes O(2)$ symmetry.

\newcounter{obrazek}

\begin{figure}[t]
\refstepcounter{obrazek}
\label{fig1}
\vspace{.2in}
\ifpdf
\centerline{\includegraphics[width=2.9in]{grid.pdf}}
\else
\centerline{\includegraphics[width=2.9in]{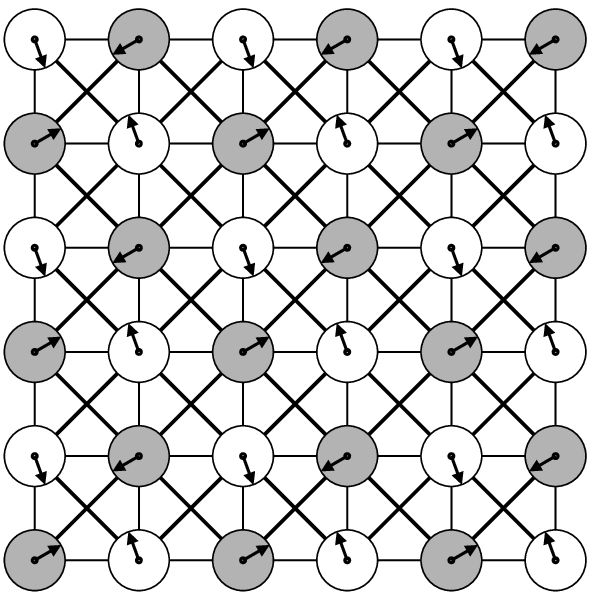}}
\fi
\vspace{.2in}
\bigskip
\begin{quote}
\fontsize{9}{5}\selectfont
{\sc Figure~\theobrazek.\ }
An example of the ground state of the Hamiltonian \eqref{Ham} on a finite grid. Here both sublattices exhibit Ne\'el state with spins alternating between 30$^\circ$ and 210$^\circ$ on one sublattice and between~110$^\circ$ and 290$^\circ$ on the other. Any other ground state can be obtained by an independent rotation of all spins in each sublattice. 
\normalsize
\end{quote}
\end{figure}

For convenience we will regard the first factor of $O(2)\otimes O(2)$ as acting on all spins and the second as acting on the \emph{relative orientations} of (the spins on) the two sublattices. 
The upshot of this work (precise theorems will be stated in Section~\ref{sec2.2}) is that, at small but positive temperatures, the order parameter space is reduced to~$\Z_2$. Although the first~$O(2)$ is restored as required by the Mermin-Wagner theorem, the remaining~$\Z_2$ is a remnant of the second~$O(2)$. Consequently, at low temperatures, there are two Gibbs states: one where there is near alignment between nearest-neighbor spins in every lattice column and the other featuring a similar alignment in every lattice row. So the continuous $O(2)\otimes O(2)$ symmetry is evidently broken; we have Gibbs state in which all that acts is the single~$O(2)$ factor. And all of this in two dimensions! 

Having arranged for the requisite two moments via procrastination, we will now reveal why this does not violate the Mermin-Wagner theorem. The answer is that the enhanced $O(2)\otimes O(2)$ symmetry was never a symmetry of the \emph{Hamiltonian}---this is both the hypothesis and the driving force of the derivations of the Mermin-Wagner theorem. Indeed, the large symmetry was only a symmetry of the ground state space and as such there is no \emph{a priori} reason to expect its persistence at finite temperatures. So everything is all right. To further confuse matters, let us remark that although the ``$\Z_2$ remnant''---the one that does get broken---was not an internal symmetry of the Hamiltonian, it is, somehow, more organic than the $O(2)$ group that contained it. This particular~$\Z_2$ may be interpreted as the natural enactor of one of the lattice symmetries (here a 90$^\circ$-rotation) which are typically associated with antiferromagnets.

The last observation is supported by the fact that there is an order parameter associated with the above phase transition. Indeed, consider the object
\begin{equation}
\label{orderpar}
\fraktura{n}_{\br}=(\bS_{\br+\hate_x}-\bS_{\br+\hate_y})\cdot\bS_{\br}
\end{equation}
whose expectation is zero at sufficiently high temperatures and non-zero (in appropriate states) at low temperatures. (In another context, this sort of symmetry breaking has been referred to as Ising nematic ordering~\cite{Abanov,Kivelson-Fradkin-Emery}.) To summarize (in case all of this has been confusing), here we have a \emph{true} long-range order but we avoid conflict with the Mermin-Wagner theorem because the $O(2)\otimes O(2)$-symmetry was never a \emph{true} symmetry of the model.

\subsection{Order by disorder}
\label{sec1.3}\noindent
In accordance with the title, the mechanism behind this ordering is called ``order by disorder''
(or, in the older vernacular, ``ordering due to disorder'').  This concept is, as of late, extremely prevalent in the physics literature; most of the recent work concerns quantum large-$S$ systems where finite~$S$ plays the role of thermal fluctuations, but the origin of this technique can be traced to the study of classical systems, see~\cite{Shender,Villain} and \cite{Henley}. In particular, in the latter reference, it is exactly the present model that was studied and this has since been referred to as the canonical model of order by disorder. 
The key words are ``spin waves'' and ``stabilization by finite temperature excitations,'' neither of which should be unfamiliar to the mathematical physicist but which, until recently, have not been exploited in tandem.

Let us proceed with the key ideas; we will attend to the obligatory citations later. For ease of exposition, let us imagine that somehow even at finite temperatures the two sublattices remain locked in their Ne\'el states. Thus there is an angle,~$\phi^\star$, which measures the relative orientation of the states on the two sublattices. Next we perform a \emph{spin-wave} calculation to account for the thermal perturbations about the ground state with fixed~$\phi^\star$. Although said instructions may have profound implications in other contexts, for present purposes this simply means ``pitch out all interactions beyond quadratic order and perform the resulting Gaussian integral.'' The upshot of such a calculation is a quantity, the \emph{spin-wave free energy}, which should then be minimized as a function of~$\phi^\star$. As we will see this minimum occurs exactly when the states are either horizontally or vertically~aligned, i.e.,~$\phi^\star=0^\circ$ or~$\phi^\star=180^\circ$.

The reader may question the moral grounds for the working assumption of finite temperature Ne\'el order which is the apparent basis of the spin-wave calculation. Of course, the cheap way out---the final arbitrator---is the fact that herein is a rigorous proof.
However, the spin-wave conclusions are not so difficult to understand. Foremost, we reemphasize that the outcome is decided purely on the basis of free energetics. A cursory examination of the calculational mechanics then reveals that in fact only two ingredients are really needed. The first is that Ne\'el order is present locally---which is certainly true at very low temperatures. The second boils down to the statement that the \emph{thermodynamic} properties in these sorts of magnets are unaffected---to first approximation---if the system is restricted to configurations that have magnetic order. In particular, the long wave-length excitations which are ultimately responsible for the break-up of ordering in two dimensions contribute insignificantly to the free energy.

Now let us discuss the historical perspective of the present paper.
The first phase in understanding this sort of problems is coming to terms with the degeneracy of the ground-state space. When these situations arise, there is a selection at finite temperature according to the ability that each state has to harbor excitations. The simplest cases, namely a finite number of ground states and a small effective activity (e.g., a large ``mass'') for the excitations have been understood by physicists for a long time and are now the subject of essentially complete mathematical theorems~\cite{PSa,PSb,Z1}. Many interesting situations with infinitely many ground states were introduced in late 1970s and early 1980s, see, e.g., \cite{Villain,Fisher-Selke}. Here intricate and/or mysterious calculations are invoked to resolve the degeneracies---often resulting in phantasmagorical phase diagrams, see e.g.~\cite{Fisher-Szpilka}---but the upshot in these situations is pretty much the same. In particular, with excruciating effort, some cases can now be proclaimed as theorems~\cite{Dinaburg-Sinai,Bricmont-Slawny}. However, the cornerstone of any systematic analysis (either mathematical or physical) is the existence of a substantial gap in the energy spectrum separating those excitations which resolve the ground-state degeneracy from the excitations that are readily available to all ground states.

The degenerate ground-state problems look very different for the classical~$O(n)$-spin models. Indeed, the continuous nature of the spins in combinations with their internal degrees of freedom almost inevitably lead to a gapless excitation spectrum. Although this sounds a lot harder, the necessary computations turn out to be far more palatable. To our knowledge, the first such example, studied in \cite{Shender}, was a frustrated FCC antiferromagnet. The system is quite similar to the one discussed here but with the ordering caused, mostly, by quantum effects. In~\cite{Henley}, studying exactly the model in \eqref{Ham}, it was demonstrated that these techniques also apply to classical systems. In the present work we will transform these classical finite-temperature derivations into a mathematical theorem. The proofs are quite tractable; all that is really required are some error estimates for the Gaussian approximations and a straightforward contour argument. To ease our way through the latter we will employ the method of chessboard estimates. In some concurrent work \cite{BCN1,BCN2}, a similar analysis is used to resolve some controversies concerning models of transition-metal oxides. However, in these ``TMO-problems,'' the ground-state spaces have additional intricacies so the beauty and simplicity of the method~is~obscured.

To make our historical perspective complete let us also relate to the existing mathematical work on systems with continuous spins. A general approach to continuous spins with degeneracies has been developed in~\cite{Dobrushin-Zahradnik,Zahradnik}. Here the method of resolution appears to be not terribly dissimilar to ours; e.g., there are quadratic approximations, Gaussian integrals, error estimates, etc. However, only a finite number of ground states are considered and we suspect that a detailed look at the ``curvature conditions'' will reveal that again there is a substantial mass gap in the excitation spectrum. Finally, from an earlier era, there are the methods based on infrared bounds~\cite{FSS,DLS,FILS1,FILS2}.
However, the reflection symmetries required to get these arguments started do not seem to hold in the system defined by~\eqref{Ham}. And even if they did, due to the infrared divergence, this would only provide misleading evidence---\emph{a la} Mermin-Wagner---that the model under consideration has no phase transition.

\section{Main results}
\label{sec2}\vspace{-3mm}
\subsection{Phase coexistence}
\label{sec2.2}\noindent
To state our results on phase coexistence in the model under consideration, we will first recall the concept of infinite-volume Gibbs measures. We begin with finite-volume counterparts thereof, also known as Gibbs specifications. Let~$\bS=(\bS_\Lambda,\bS_{\Lambda^\cc})$ be a spin configuration where~$\bS_\Lambda$ and~$\bS_{\Lambda^\cc}$ denote the corresponding restrictions to~$\Lambda$ and~$\Lambda^\cc$, respectively. Let~$\mathscr{H}_\Lambda(\bS_\Lambda,\bS_{\Lambda^\cc})$ be the restriction of \eqref{Ham} to pairs of sites at least one of which is in~$\Lambda$. Then we let~$\mu_\Lambda^{(\bS_{\Lambda^\cc})}$ be the measure on configurations in~$\Lambda$ defined by
\begin{equation}
\label{fin-vol-GM}
\mu_\Lambda^{(\bS_{\Lambda^\cc})}(\textd\bS_\Lambda)=
\frac{e^{-\beta\mathscr{H}_\Lambda(\bS_\Lambda,\bS_{\Lambda^\cc})}}
{Z_\Lambda(\bS_{\Lambda^\cc})}\Omega_\Lambda(\textd\bS_\Lambda).
\end{equation}
Here~$\Omega_\Lambda$ denotes the product Lebesgue measure on the unit circle, one for each~$\br\in\Lambda$. Following the ``DLR-philosophy,'' see~\cite{Georgii}, the infinite-volume Gibbs measures are those measures on full configurations on~$\Z^2$ whose conditional probability in a finite volume~$\Lambda$ given the configuration in the complement is exactly the object in \eqref{fin-vol-GM}. 

In accord with the standard terminology, see \cite{Georgii}, we will say that there is a \emph{phase coexistence} for parameters~$J$,~$\gamma$ and~$\beta$ if there exists more than one infinite-volume Gibbs measure for the interaction \eqref{Ham} and inverse temperature~$\beta$. To adhere with mathematical-physics notation, we will refer to the Gibbs measures as \emph{Gibbs states} and we will denote the expectations with respect to such states by symbol~$\langle-\rangle_\beta$.

\smallskip
Now we are in a position to state the main result of this paper.

\begin{theorem}
\label{thm2.1}
Consider the model as defined above with fixed~$J\in(0,\infty)$ and~$\gamma\in(0,2)$. Then there exists a~$\beta_0\in(0,\infty)$ and a function~$\beta\mapsto\epsilon(\beta)$ satisfying~$\epsilon(\beta)\to0$ as~$\beta\to\infty$ such that the following holds: For each~$\beta\ge\beta_0$ there exist two distinct Gibbs states~$\langle-\rangle^{(x)}_\beta$ and~$\langle-\rangle^{(y)}_\beta$ such~that
\begin{equation}
\label{2.2}
\bigl|\langle\bS_{\br}\cdot\bS_{\br'}\rangle^{(\alpha)}_\beta+1\bigr|\le\epsilon(\beta)
\end{equation}
whenever~$\br,\br'$ are next-nearest neighbors in~$\Z^2$, and
\begin{equation}
\label{2.3}
\bigl|\langle\bS_{\br}\cdot\bS_{\br'}\rangle^{(\alpha)}_\beta-1\bigr|\le\epsilon(\beta)
\end{equation}
whenever~$\br,\br'\in\Z^2$ are such that $\br'=\br+\hate_\alpha$.
\end{theorem}

Let us informally describe the previous result. First, on both even and odd sublattice of~$\Z^2$
we have a (local) antiferromagnetic order. The distinction between the two states is that in $\langle-\rangle^{(x)}_\beta$ the nearest-neighbor spins on~$\Z^2$ are aligned in the~$x$ direction and antialigned in the~$y$ direction, while in $\langle-\rangle^{(y)}_\beta$ the two alignment directions are interchanged. In particular, it is clear that the order parameter~$\fraktura{n}_{\br}$, defined in \eqref{orderpar}, has positive expectation in the~$x$-state $\langle-\rangle^{(x)}_\beta$ and negative expectation in the~$y$-state $\langle-\rangle^{(y)}_\beta$.
Since, as mentioned previously, Gibbsian uniqueness guarantees that~$\langle\fraktura{n}_{\br}\rangle_\beta=0$ at sufficiently high temperatures, we have a \emph{bone fide} phase transition of the ``usual'' type.

\smallskip
Despite the existence of multiple low-temperature Gibbs states, we emphasize that no claim has been made about the actual direction that the spins will be aligned to. On the contrary, we have the following easy corollary of the aforementioned Mermin-Wagner theorem:

\begin{theorem}
\label{thm2.2}
Consider the model as defined above with~$J,\gamma\in\R$ fixed and let~$\langle-\rangle_\beta$ be any infinite-volume Gibbs state at inverse temperature~$\beta$. Then~$\langle-\rangle_\beta$ is invariant under the simultaneous rotation of all spins and, in particular,~$\langle\bS_{\br}\rangle_\beta=\bzero$ for all~$\br\in\Z^2$.
\end{theorem}

The authors do not see any significant obstruction of Theorem~\ref{thm2.1} (appropriately modified) in the cases~$n>2$ and~$d>2$. For the case under consideration, namely,~$n=2$ and~$d=2$, it may be presumed that there is a slow decay of correlations at sufficiently low temperatures. Here it is conceivable that, with great effort, this could be proved on the basis of technology that is currently available~\cite{Frohlich-Spencer,Marchetti-Klein,Dimock-Hurd}. The anticipation is that for~$d\ge3$ and~$n\ge2$ there are actual sublattice Ne\'el states while for~$d=2$ and~$n>2$ the decorrelations should be exponential. However, we do not expect to see a proof of any statement along these lines in the near future.

\subsection{Outline of the proof}
We proceed by an informal outline of the proof of our main result (Theorem~\ref{thm2.1}). The argument hinges on the following three observations:
\settowidth{\leftmargini}{(11)}
\begin{enumerate}
\item[(1)]
Suppose~$\Delta$ is a number that satisfies
\begin{equation}
\label{2.4}
\beta J\Delta^2\gg1.
\end{equation}
Then the (angular) difference of any typical pair of \emph{next}-nearest neighbor spins will not deviate by more than~$\Delta$ from the energetically optimal configuration.
\item[(2)]
In situations when~(1) applies and under the additional assumption that~$\Delta$ also satisfies
\begin{equation}
\label{2.5}
\beta J\Delta^3\ll1,
\end{equation}
then all important contributions to the \emph{free} energy of the system will come from a quadratic---or \emph{spin-wave}---approximation to the Hamiltonian.
\item[(3)]
Finally, if~$F(\phi^\star)$ denotes the spin-wave free energy above the ground state where one sublattice is rotated by angle~$\phi^\star$ relative to the other (see Fig.~\ref{fig1}), then~$F(\phi^\star)$ is minimized only at~$\phi^\star=0^\circ$ or~$\phi^\star=180^\circ$.
\end{enumerate}
(The mathematical statements corresponding to (1-3) above are formulated as Theorems~\ref{thm3.2} and~\ref{thm3.3} in Section~\ref{sec3.1}.) We observe that the necessary~$\Delta$ as stipulated by \twoeqref{2.4}{2.5} defines a running scale---not too big and not too small---which obviously tends to zero as~$\beta\to\infty$.

Here is how these observations will be combined together to establish long-range order: We partition the lattice in blocks of side~$B$. On the basis of~(1) above, every block will with high probability exhibit a near ground-state configuration, which by~(2-3) will have the sublattices either nearly aligned or nearly antialigned. Then we need to show that each of the two possibilities are stable throughout the entire system. For that we will resort to a standard Peierls' argument. Here the crucial observation (see Lemma~\ref{lemma4.6}) is that two ``good'' blocks with different type of alignment between sublattices are necessarily separated by a ``surface'' of ``bad'' blocks---that is those which either contain energetically charged pair of nearest-neighbor spins or whose spin-wave free energy exceeds the absolute minimum by a positive amount. 

Appealing to chessboard estimates (see Section~\ref{sec4.1}), the probability of a particular ``surface'' can be factorized---as a bound---into the product over the constituting blocks. It turns out that the energetically frustrated ``bad'' blocks are suppressed once
\begin{equation}
\label{2.6}
\beta J\Delta^2\gg\log B,
\end{equation}
while the entropically frustrated blocks are suppressed once the excess spin-wave free energy \emph{times~$B^2$} is sufficiently large. Under the conditions \twoeqref{2.5}{2.6} and~$B\gg1$ the entropy of the above ``surfaces'' can be controlled. The desired phase coexistence then follows by standard arguments. 

A couple of remarks are in order: Due to the perfect scaling properties of Gaussian distributions the suppression extracted from the spin-wave calculation is \emph{independent of~$\beta$}---the desired decay is achieved solely by choosing~$B$ sufficiently large. Large~$\beta$ is needed only to suppress large deviations away from the ``perfect'' ground states. Notwithstanding, for (technical) ease of exposition we will have to make~$B$ increase slowly with~$\beta$; see \eqref{DBass} for the precise relation of~$\Delta$,~$B$ and~$\beta$.

\smallskip
The various steps of the proof are laid out in the following order: In Section~\ref{sec3} we carry out the harmonic approximation and provide the needed control of the spin-wave free energy. In Section~\ref{sec4} we invoke chessboard estimates and some straightforward bounds to control the contour expansion. The actual proof of Theorem~\ref{thm2.1} comes in Section~\ref{sec4.3}.

\section{Spin-wave calculations}
\label{sec3}\noindent
As mentioned above, the underpinning of our proof of the main result is (the outcome of) a spin-wave free-energy calculation.
This calculation involves simply working with the harmonic approximation of the Hamiltonian \eqref{Ham} for deviations away from a fixed ground state. The calculation itself is straightforward although special attention must be paid to the ``zero mode.'' 
For reasons that will become clear in Section~\ref{sec4}---and also to make discrete Fourier transform readily available---all of the derivations in this section will be carried out on the lattice torus~$\T_L$ of~$L\times L$-sites. Here, for technical convenience, we will restrict~$L$ to multiples of four so that we can assure an equal status of the two Ne\'el states.

\subsection{Harmonic approximation}
\label{sec3.1}\noindent
We will begin by an explicit definition of the torus Hamiltonian. Here and henceforth we will parametrize the spins by angular variables~$\btheta=(\theta_{\br})$ which are related to the~$\bS_{\br}$'s by the usual expression $\bS_{\br}=(\cos\theta_{\br},\sin\theta_{\br})$. (Of course, the~$\theta_{\br}$'s are always to be interpreted only modulo~$2\pi$.) Up to irrelevant constants, the corresponding torus Hamiltonian~$\mathscr{H}_L$ can then be written as
\begin{multline}
\label{3.1}
\qquad
\mathscr{H}_L(\btheta)=J\sum_{\br\in\T_L}
\bigl\{2+\cos(\theta_{\br}-\theta_{\br+\hate_x+\hate_y})
+\cos(\theta_{\br}-\theta_{\br+\hate_x-\hate_y})
\bigr\}
\\+J\gamma\sum_{\br\in\T_L}
\bigl\{\cos(\theta_{\br}-\theta_{\br+\hate_x})
+\cos(\theta_{\br}-\theta_{\br+\hate_y})\bigr\}.
\qquad
\end{multline}
The spin-wave calculations are only meaningful in the situations where each of the sublattices is more or less aligned with a particular Ne\'el state. To describe the overall and relative orientation of the spins on the even and odd sublattices we will need two angles~$\theta^\star$ and~$\phi^\star$, respectively. Depending on the parities of the coordinates of~$\br$, we will write the~$\theta_{\br}$ for~$\br=(x,y)$ in terms of the \emph{deviation variables}~$\vartheta_{\br}$ as follows:
\begin{equation}
\label{varthdef}
\theta_{\br}=\vartheta_{\br}+\begin{cases}
\theta^\star,\qquad&x,y\text{-even},
\\
\theta^\star+\phi^\star,\qquad&x\text{-odd, }y\text{-even},
\\
\theta^\star+\pi,\qquad&x,y\text{-odd},
\\
\theta^\star+\phi^\star+\pi,\qquad&x\text{-even, }y\text{-odd}.
\end{cases}
\end{equation}
Obviously, only the relative angle~$\phi^\star$ will appear in physically relevant quantities; the overall orientation~$\theta^\star$ simply factors out from all forthcoming expressions.

The principal object of interest in this section is the finite-volume free energy, which will play an important role in the estimates of ``entropically-disfavored'' block events in Section~\ref{sec4}. For reasons that will become clear later, we will define this quantity by the formula
\begin{equation}
\label{3.7}
F_{L,\Delta}(\phi^\star)=-\frac1{L^2}\log\int e^{-\beta\mathscr{H}_L(\btheta)}\chi_{L,\Delta}(\btheta)\,
\Bigl(\frac{\beta J}{2\pi}\Bigr)^{L^2/2} 
\prod_{\br\in\T_L}\textd\theta_{\br}.
\end{equation}
Here~$\textd\theta_{\br}$ is the Lebesgue measure on unit circle and~$\chi_{L,\Delta}(\btheta)=\chi_{L,\Delta}(\btheta;\phi^\star,\theta^\star)$ is the indicator that the deviation quantities~$\bvartheta$, defined from~$\btheta$ as detailed in \eqref{varthdef}, satisfy $|\vartheta_{\br}|<\Delta$ for all~$\br\in\T_L$. 
The factors of~$\tfrac{\beta J}{2\pi}$ have been added for later convenience.

\smallskip
The goal of this section is to (approximately) evaluate the thermodynamic limit of the quantity~$F_{L,\Delta}(\phi^\star)$ and characterize where it achieves its minima.
As is standard in heuristic calculations of this sort, we will first replace the Hamiltonian \eqref{3.1} by its appropriate quadratic approximation. We will express the resulting quantity directly in variables~$\vartheta_{\br}$:
\begin{multline}
\qquad
\mathscr{I}_{L,\phi^\star}(\bvartheta)=\frac{\beta J}2\sum_{\br\in\T_L}
\bigl\{(\vartheta_{\br}-\vartheta_{\br+\hate_x+\hate_y})^2
+(\vartheta_{\br}-\vartheta_{\br+\hate_x-\hate_y})^2
\bigr\}
\\+\frac{\beta J}2\gamma\cos(\phi^\star)\sum_{\br\in\T_L}
\bigl\{(\vartheta_{\br}-\vartheta_{\br+\hate_x})^2
+(\vartheta_{\br}-\vartheta_{\br+\hate_y})^2\bigr\}.
\qquad
\end{multline}
This approximation turns the integral in \eqref{3.7} into a Gaussian integral. As we will see later, here the indicator in \eqref{3.7} can be handled in terms of upper and lower bounds which allow ``diagonalization'' of the covariance matrix by means of Fourier variables. The result, expressed in the limit~$L\to\infty$, is the following momentum integral:
\begin{equation}
\label{3.8}
F(\phi^\star)=\frac12\int_{[-\pi,\pi]^2}\frac{\textd\bk}{(2\pi)^2}\log D_{\bk}(\phi^\star),
\end{equation}
where
\begin{equation}
\label{3.9}
D_{\bk}(\phi^\star)=|1-e^{\texti(k_1+k_2)}|^2+|1-e^{\texti(k_1-k_2)}|^2+\gamma\cos(\phi^\star)\bigl(|1-e^{\texti k_1}|^2-|1-e^{\texti k_1}|^2\bigr).
\end{equation}
Here~$k_1$ and~$k_2$ are the Cartesian components of vector~$\bk$. The quantity~$F(\phi^\star)$ has the interpre\-tation---justified via the preceding derivation---as the \emph{spin-wave free energy}. As is checked by direct calculation, for $\gamma\in[0,2)$ we have~$D_{\bk}(\phi^\star)>0$ almost surely with respect to the (normalized) Lebesgue measure on $[-\pi,\pi]^2$.

\smallskip
Having sketched the main strategy and defined the relevant quantities, we can now pass to the statements of (admittedly dry) mathematical theorems. First, we express the conditions under which the above approximate calculation can be performed:

\begin{theorem}
\label{thm3.2}
Given~$\epsilon>0$ and $\gamma\in[0,2)$, there exists~$\delta=\delta(\epsilon,\gamma)>0$ such that if~$\beta J$,~$\Delta$ and~$\delta$ satisfy the bounds
\begin{equation}
\label{3.10}
\beta J\Delta^3\le\delta
\quad\text{and}\quad
\beta J\Delta^2\ge1/\delta,
\end{equation}
then
\begin{equation}
\label{3.11}
\limsup_{L\to\infty}\bigl|F_{L,\Delta}(\phi^\star)-F(\phi^\star)\bigr|\le\epsilon
\end{equation}
holds for every~$\phi^\star\in(-\pi,\pi]$.
\end{theorem}

The proof is postponed to Section~\ref{sec3.2}.
Having demonstrated the physical meaning of the function~$\phi^\star\mapsto F(\phi^\star)$, we can now characterize its absolute minimizers:

\begin{theorem}
\label{thm3.3}
For all $\gamma\in(0,2)$, the absolute minima of function~$\phi^\star\mapsto F(\phi^\star)$ occur (only) at the points~$\phi^\star=0^\circ$ and~$\phi^\star=180^\circ$.
\end{theorem}

\begin{proofsect}{Proof}
The proof is an easy application of Jensen's inequality. Indeed, let~$a\in[0,1]$ be the number such that $2a-1=\cos(\phi^\star)$. Then we can write
\begin{equation}
D_{\bk}(\phi^\star)=a D_{\bk}(0^\circ)+(1-a)D_{\bk}(180^\circ).
\end{equation}
Since~$D_{\bk}(0^\circ)$ is not equal to~$D_{\bk}(180^\circ)$ almost surely with respect to~$\textd\bk$ (this is where we need that~$\gamma>0$), the concavity of the logarithm and Jensen's inequality imply that~$F(\phi^\star)>aF(0^\circ)+(1-a)F(180^\circ)$ whenever $a\ne0,1$. This shows that the only absolute minima that~$F$ can have are~$0^\circ$ and~$180^\circ$. Now~$F$ is continuous (under the assumption that~$\gamma<2$) and periodic, and so there exists at least one point in~$(-\pi,\pi]$ where it attains its absolute minimum. But~$F(0^\circ)=F(180^\circ)$ and so~$\phi^\star\mapsto F(\phi^\star)$ is minimized by both~$\phi^\star=0^\circ$ and $\phi^\star=180^\circ$.
\end{proofsect}

\subsection{Proof of Theorem~\ref{thm3.2}}
\label{sec3.2}\noindent
Throughout the proof we will fix~$J\in(0,\infty)$ and~$\gamma\in[0,2)$ and suppress these from our notation whenever possible.
Since everything is founded on harmonic approximation of the Hamiltonian, the starting point is some control of the error that this incurs:

\begin{lemma}
\label{lemma3.1}
There exists a constant~$c_1\in(0,\infty)$ such that the following holds:
For any $\Delta\in(0,\infty)$, any~$\theta^\star,\phi^\star\in(-\pi,\pi]$ and any configuration ~$\btheta=(\theta_{\br})$ of angle variables on~$\T_L$, if the corresponding~$\bvartheta=(\vartheta_{\br})$ satisfy $|\vartheta_{\br}|<\Delta$ for all  $\br\in\T_L$, then
\begin{equation}
\label{HIbound}
\bigl|\beta\mathscr{H}_L(\btheta)-\mathscr{I}_{L,\phi^\star}(\bvartheta)\bigr|<c_1(1+\gamma)\beta J\Delta^3L^2.
\end{equation}
\end{lemma}

\begin{proofsect}{Proof}
We begin by noting that~$|\vartheta_{\br}|<\Delta$ for all~$\br\in\T_L$ implies that~$|\vartheta_{\br}-\vartheta_{\br'}|<2\Delta$ for all pairs of nearest and next-nearest neighbors~$\br,\br'\in\T_L$. This and the uniform bound
\begin{equation}
\Bigl|\,\cos(a+x)-\bigl(\cos(a)-\sin(a)x-\tfrac12\cos(a)x^2\bigr)\Bigr|\le\frac{|x|^3}6,
\end{equation}
show that, at the cost of an error as displayed in \eqref{HIbound}, we can replace all trigonometric factors in \eqref{3.1} by their second-order Taylor expansion in differences of~$\vartheta_{\br}$. Hence, we just need to show that these Taylor polynomials combine into the expression for~$\mathscr{I}_{L,\phi^\star}$.

It is easily checked that the zeroth order Taylor expansion in~$\vartheta_{\br}$ exactly vanishes. This is a consequence of the fact that for~$\bvartheta\equiv0$ we are in a ground state where, as argued before, both sublattices can be independently rotated. This means we can suppose that~$\theta^\star=\phi^\star=0^\circ$ in \eqref{3.1} at which point it is straightforward to verify that~$\mathscr{H}_L(\btheta)$ actually vanishes.
Similarly easy it is to verify that the quadratic terms yield exactly the expression for~$\mathscr{I}_{L,\phi^\star}$.
It thus remains to prove that there are no linear terms in~$\vartheta_{\br}$'s. 

First we will note that all next-nearest neighbor terms in the Hamiltonian certainly have this property because there we have~$\theta_{\br}-\theta_{\br+\hate_x\pm\hate_y}\approx0$ or~$\pi$, at which points the derivative of the cosine vanishes. Hence we only need to focus on the nearest-neighbor part of the Hamiltonian---namely, the second sum in \eqref{3.1}---which we will temporarily denote by~$\mathscr{H}^{\text{nn}}_L$. Here we will simply calculate the derivative of~$\mathscr{H}^{\text{nn}}_L$ with respect to~$\vartheta_{\br}$:
\begin{multline}
\qquad
\frac\partial{\partial\vartheta_{\br}}\mathscr{H}^{\text{nn}}_L(\btheta)
\Bigl|_{\bvartheta\equiv\bzero}=
\sin(\theta_{\br+\hate_x}-\theta_{\br})
+\sin(\theta_{\br+\hate_y}-\theta_{\br})
\\-\bigl\{\sin(\theta_{\br}-\theta_{\br-\hate_x})
+\sin(\theta_{\br}-\theta_{\br-\hate_y})\bigr\},
\qquad
\end{multline}
where the~$\theta_{\br}$ on the right-hand side should be set to the ``ground-state'' values.
To make the discussion more explicit, suppose that~$\br$ has both coordinates even. Then an inspection of \eqref{varthdef} shows that the first sine is simply~$\sin(\phi^\star)$ while the second sine evaluates to~$\sin(\phi^\star+\pi)=-\sin(\phi^\star)$. The net contribution of these two terms is thus zero. Similarly, the third and the fourth sine also cancel out. The other possibilities for~$\br$ are handled analogously.
\end{proofsect}

Using the harmonic approximation of the Hamiltonian, let us now consider the corresponding Gaussian equivalent of the integral in \eqref{3.7}:
\begin{equation}
\label{3.12}
Q_{L,\Delta}(\phi^\star)=\int e^{-\mathscr{I}_{L,\phi^\star}(\bvartheta)}
\widetilde\chi_{L,\Delta}(\bvartheta)\,
\Bigl(\frac{\beta J}{2\pi}\Bigr)^{L^2/2}
\prod_{\br\in\T_L}\textd\vartheta_{\br},
\end{equation}
where~$\textd\vartheta_{\br}$ is the Lebesgue measure on~$\R$ and $\widetilde\chi_{L,\Delta}(\bvartheta)$ is the indicator that $|\vartheta_\br|<\Delta$ for all $\br\in\T_L$.
Our next goal is to evaluate the effect of this indicator, which we will accomplish by proving an upper and lower bound on~$Q_{L,\Delta}(\phi^\star)$. We commence with the easier of the two, the upper bound:

\begin{lemma}
\label{lemma3.3}
For all~$\beta\in(0,\infty)$, all~$\Delta\in(0,\infty)$ and all~$\phi^\star\in(-\pi,\pi]$,
\begin{equation}
\limsup_{L\to\infty}\frac{\log Q_{L,\Delta}(\phi^\star)}{L^2}\le -F(\phi^\star).
\end{equation}
\end{lemma}

\begin{proofsect}{Proof}
The argument is relatively straightforward so we will be correspondingly brief. (A more verbose argument along these lines can be found in \cite{BCN1}.) Pick a~$\lambda>0$. We will invoke the exponential Chebyshev inequality in the form
\begin{equation}
\widetilde\chi_{L,\Delta}(\bvartheta)\le
e^{\frac12\beta J\lambda\Delta L^2}\exp\Bigl\{-\frac12\beta J\lambda\sum_{\br\in\T_L}|\vartheta_{\br}|^2\Bigr\}.
\end{equation}
Next we plug this bound into \eqref{3.12}, diagonalize~$\mathscr{I}_{L,\phi^\star}$ by passing to the Fourier components $\widehat\vartheta_{\bk}=L^{-1}
\sum_{\br\in\T_L}\vartheta_{\br}e^{\texti\br\cdot\bk}$ and perform the Gaussian integrals with the result
\begin{equation}
Q_{L,\Delta}(\phi^\star)\le e^{\frac12\beta J\lambda\Delta L^2}
\prod_{\bk\in\T_L^\star}\frac1{[\lambda+D_{\bk}(\phi^\star)]^{1/2}}.
\end{equation}
Here~$\T_L^\star=\{2\pi L^{-1}(n_1,n_2)\colon n_i=1,2,\dots,L\}$ is the reciprocal lattice and~$D_{\bk}(\phi^\star)$ is as defined in \eqref{3.9}. The result now follows by taking logarithm, dividing by~$L^2$ and invoking the limits~$L\to\infty$ followed by~$\lambda\downarrow0$---with the last limit justified by the Monotone Convergence Theorem.
\end{proofsect}

The corresponding lower bound is then stated as follows:

\begin{lemma}
\label{lemma3.4}
For all~$\beta\in(0,\infty)$, all~$\Delta\in(0,\infty)$, all~$\phi^\star\in(-\pi,\pi]$ and all~$\lambda>0$ satisfying $\beta J\Delta^2\lambda>1$, we have
\begin{equation}
\liminf_{L\to\infty}\frac{\log Q_{L,\Delta}(\phi^\star)}{L^2}\ge -F(\phi^\star,\lambda)+\log\Bigl(1-\frac1{\beta J\Delta^2\lambda}\Bigr),
\end{equation}
where~$F(\phi^\star,\lambda)$ is given by the same integral as in \eqref{3.8} with~$D_{\bk}(\phi^\star)$ replaced by $\lambda+D_{\bk}(\phi^\star)$.
\end{lemma}

\begin{proofsect}{Proof}
Again, we will be fairly succinct. Let~$\lambda>0$. We begin by considering the Gaussian measure defined by
\begin{equation}
\label{Gmu}
\BbbP_\lambda(\textd\bvartheta)=\frac1{Q_L(\phi^\star,\lambda)}\exp\Bigl\{-\mathscr{I}_{L,\phi^\star}(\bvartheta)-\frac12\beta J\lambda\sum_{\br\in\T_L}|\vartheta_{\br}|^2\Bigr\}
\Bigl(\frac{\beta J}{2\pi}\Bigr)^{L^2/2}
\prod_{\br\in\T_L}\textd\vartheta_{\br}
\end{equation}
where~$Q_L(\phi^\star,\lambda)$ is an appropriate normalization constant. Let~$\E_\lambda$ denote the corresponding expectation. A simple bound shows that we have
\begin{equation}
\label{3.18}
Q_{L,\Delta}(\phi^\star)\ge Q_L(\phi^\star,\lambda)\E_\lambda(\widetilde\chi_{\Delta,L}),
\end{equation}
which reduces the desired estimates to two items: a calculation of the integral~$Q_L(\phi^\star,\lambda)$ and a lower bound on $\E_\lambda(\widetilde\chi_{\Delta,L})$.

The first problem on the list is dispensed with similarly as in the proof of Lemma~\ref{lemma3.3}, so we just state the result:
\begin{equation}
\label{3.19}
\lim_{L\to\infty}\frac{\log Q_L(\phi^\star,\lambda)}{L^2}=-F(\phi^\star,\lambda).
\end{equation}
As far as the second item on the list is concerned, here we use that by the results of \cite{BCG}  the magnitudes of the Gaussian field with distribution \eqref{Gmu} are positively correlated. 
(An alternative proof of this fact uses reflection positivity.) Invoking the product structure of~$\widetilde\chi_{\Delta,L}$ and translation invariance of~$\BbbP_\lambda$, we thus have
\begin{equation}
\label{3.20}
\E_\lambda(\widetilde\chi_{\Delta,L})\ge\BbbP_\lambda\bigl(|\vartheta_0|<\Delta\bigr)^{L^2},
\end{equation}
where~$\vartheta_0$ is the variable at the origin of the torus.

It remains to bound $\BbbP_\lambda(|\vartheta_0|<\Delta)$ from below, which we will do by estimating the complementary probability from above. We will pass to the Fourier components~$\widehat\vartheta_{\bk}$ defined as in the proof of Lemma~\ref{lemma3.3}. Under the measure \eqref{Gmu}, these components have zero mean, the random variables~$\widehat\vartheta_{\bk}$ and~$\widehat\vartheta_{\bk'}^*$ for different~$\bk$ and~$\bk'$ are uncorrelated (a consequence of translation invariance), while for the autocorrelation function we get
\begin{equation}
\E_\lambda\bigl(|\widehat\vartheta_{\bk}|^2\bigr)=\frac1{\beta J}\,
\frac1{\lambda+D_{\bk}(\phi^\star)}\le\frac1{\beta J\lambda}.
\end{equation}
This allows us to use the (quadratic) Chebyshev inequality to derive
\begin{equation}
\BbbP_\lambda(|\vartheta_0|\ge\Delta)\le\frac{\E_\lambda(|\vartheta_0|^2)}{\Delta^2}=
\frac1{L^2}\sum_{\bk\in\T_L^\star}\frac{\E_\lambda(|\widehat\vartheta_{\bk}|^2)}{\Delta^2}
\le\frac1{\beta J\Delta^2\lambda}.
\end{equation}
Inserting this into \eqref{3.20} and applying \eqref{3.18} and \eqref{3.19}, the rest of the proof boils down to taking logs, dividing by~$L^2$ and letting~$L\to\infty$.
\end{proofsect}

Now we are ready to prove the principal approximation theorem:

\begin{proofsect}{Proof of Theorem~\ref{thm3.2}}
We just assemble together the previously discussed ingredients. First, our constraints \eqref{3.10} imply that~$\Delta\le\delta^2$ and so we can assume that~$\Delta<\pi$. Under this condition the integrals in \eqref{3.7} and \eqref{3.12} are over the same set of~$\vartheta_\br$'s and so by Lemma~\ref{lemma3.1} we have the uniform bound
\begin{equation}
\label{3.23}
\Bigl|\frac{\log Q_{L,\Delta}(\phi^\star)}{L^2}-F_{L,\Delta}(\phi^\star)\Bigr|\le c_1(1+\gamma)\beta J\Delta^3.
\end{equation}
Second, Lemmas~\ref{lemma3.3}-\ref{lemma3.4} ensure that
\begin{equation}
\label{3.24}
\limsup_{L\to\infty}\Bigl|\frac{\log Q_{L,\Delta}(\phi^\star)}{L^2}
-F(\phi^\star)\Bigr|\le \bigl|F(\phi^\star)-F(\phi^\star,\lambda)\bigr|+\log\Bigl(1-\frac1{\beta J\Delta^2\lambda}\Bigr).
\end{equation}
By the assumptions in \eqref{3.10}, given an~$\epsilon>0$ we can choose~$\delta>0$ such that the right-hand side of \eqref{3.23} is smaller than~$\epsilon/2$. On the other hand, since $F(\phi^\star,\lambda)$ increases to~$F(\phi^\star)$ as~$\lambda\downarrow0$ and since~$\beta\Delta^2\ge1/\delta$, we can certainly choose a $\lambda>0$ (satisfying $\beta J\Delta^2\lambda>1$) and adjust~$\delta$ such that also the right-hand side of \eqref{3.24} is less than~$\epsilon/2$. Combining these observations, the desired bound \eqref{3.11} is proved.
\end{proofsect}

\begin{remark}
Physically motivated readers will notice that in both Lemmas~\ref{lemma3.3} and~\ref{lemma3.4} we have introduced a ``mass'' into the spin-wave spectrum before (or while) removing the indicator~$\widetilde\chi_{L,\Delta}$. The primary reason for this is the bad behavior of the zero Fourier mode for which the ``spin-wave Hamiltonian'' $\mathscr{I}_{L,\phi^\star}$ provides no decay in the Gaussian weight.
\end{remark}

\section{Proof of phase coexistence}
\label{sec4}\noindent
Having discussed the spin-wave approximations (which will be essential for the arguments in this section), we are now ready to start with the proof of phase coexistence. Our basic tool in this section will be the chessboard estimates, so we will begin by introducing the notation needed for applications of this technique.

\subsection{Chessboard estimates}
\label{sec4.1}\noindent
As mentioned previously, in order use chessboard estimates, for technical reasons, we have to confine our technical considerations to toroidal geometries. Again we will use~$\T_L$ to denote the torus of~$L\times L$ sites (as in Section~\ref{sec3} we restrict~$L$ to multiples of four). We will consider several events which will all take place in a box~$\Lambda_B$ of~$(B+1)\times(B+1)$ sites (which, for definiteness, we will assume to be placed with its lower-left corner at the torus ``origin''). Since we want to be able to cover~$\T_L$ by translates of~$\Lambda_B$, we will assume that~$L$ is an even multiple of~$B$. Thus, if~$\AA$ is an event in~$\Lambda_B$, then its translate by~$t_1B$ lattice units in the~$x$-direction and~$t_2B$ units in the~$y$-direction will be denoted by~$\tau_{\bt}(\AA)$, where~$\bt=(t_1,t_2)$. Here~$\bt$ takes values in a factor torus, namely,~$\bt\in\T_{L/B}$. Note that events in the ``neighboring'' translates of~$\Lambda_B$ may both depend on the shared side of the corresponding boxes.

Let~$\BbbP_{L,\beta}$ denote the Gibbs measure on~$\T_L$ defined from the appropriate torus version of the Hamiltonian~\eqref{Ham} and inverse temperature~$\beta$. Specifically, using the ``spin-version'' of the Hamiltonian \eqref{3.1}, the Radon-Nikodym derivative of~$\BbbP_{L,\beta}$ with respect to the \emph{a priori} spin measure~$\Omega_{\T_L}$ is~$e^{-\beta\mathscr{H}_L(\bS)}/Z_{L,\beta}$, where~$Z_{L,\beta}$ is the corresponding partition function.
The statement of the chessboard estimates will be considerably easier if we restrict our attention to reflection symmetric events, which are those~$\AA$ for which~$\bS\in\AA$ implies that the corresponding reflection~$\bS^\star$ in any coordinate plane passing through the center of~$\Lambda_B$ satisfies~$\bS^\star\in\AA$. For these events we will also define the constrained partition function
\begin{equation}
\label{4.1}
Z_{L,\beta}(\AA)=Z_{L,\beta}\,\Bigl\langle\,
\prod_{\bt\in\T_{L/B}}\1_{\tau_{\bt}(\AA)}\Bigr\rangle_{L,\beta}.
\end{equation}
Here~$\1_{\tau_{\bt}(\AA)}$ is the indicator of~$\tau_{\bt}(\AA)$ and $\langle-\rangle_{L,\beta}$ denotes the expectation with respect to~$\BbbP_{L,\beta}$.

\smallskip
Then we have:

\begin{theorem}[Chessboard estimates]
\label{thm4.1}
Consider the Gibbs measure~$\BbbP_{L,\beta}$ as defined above. Let $\AA_1,\dots,\AA_m$ be a collection of (not necessarily distinct) reflection-symmetric events in~$\Lambda_B$ and let~$\bt_1,\dots,\bt_m$ be distinct vectors from~$\T_{L/B}$. Then
\begin{equation}
\label{CE}
\BbbP_{L,\beta}\Bigl(\,\bigcap_{j=1}^m\tau_{\bt_j}(\AA_j)\Bigr)
\le\prod_{j=1}^m \Bigl(\frac{Z_{L,\beta}(\AA_j)}{Z_{L,\beta}}\Bigr)^{(B/L)^2}.
\end{equation}
\end{theorem}

\begin{proofsect}{Proof}
This is the standard chessboard estimate implied by the reflection positivity condition \cite{FSS,FILS1,FILS2}. Here we consider reflection positivity in planes ``through'' sites, which holds in our case because we have only nearest and next-nearest neighbor interactions.
\end{proofsect}

Unfortunately, as often happens with chessboard estimates, we may not be able to estimate directly the quantity~$Z_{L,\beta}(\AA)$ for the desired event under consideration. Instead, we will decompose~$\AA$ into a collection of more elementary events for which this estimation is easier. Here chessboard estimates can be used to establish the following standard (and often implicitly used) subadditivity property:

\begin{lemma}[Subadditivity]
\label{lemma4.2}
Let the torus~$\T_L$ and the block~$\Lambda_B$ be as above and let us consider reflection-symmetric events~$\AA$ and~$(\AA_k)_{k\in\scrK}$ in~$\Lambda_B$. If~$\AA\subseteq\bigcup_{k\in\scrK}\AA_k$, then
\begin{equation}
Z_{L,\beta}(\AA)^{(B/L)^2}\le\sum_{k\in\scrK}Z_{L,\beta}(\AA_k)^{(B/L)^2}.
\end{equation}
\end{lemma}

\begin{proofsect}{Proof}
See, e.g., Lemma~6.3 in \cite{BCN1}.
\end{proofsect}

Our succinct recount of the chessboard estimates is now complete. Readers wishing to obtain more details on this and related topics are referred to (still succinct) Section~6.1 of \cite{BCN1} or the classic references~\cite{FSS,FILS1,FILS2} and~\cite{Senya2}.

\subsection{Good and bad events}
\label{sec4.2}\noindent
Here we introduce the notion of good and bad blocks and events. Roughly speaking, a block is good if all spins on both sublattices are tolerably close to a Ne\'el state and where the relative orientation of the two Ne\'el states is near one of the two optimal values predicted by the spin-wave approximation. The bad blocks will of course be all those that are not good. 
Both these notions will involve two parameters: the spin-deviation scale~$\Delta$ encountered already in Section~\ref{sec3}, and the scale~$\kappa$ marking the distance to a spin-wave minimum which is still considered good. We will keep~$\kappa$ small but fixed, while~$\Delta$ will have to be decreased (and the block scale~$B$ will have to be increased, albeit only slowly) as~$\beta$ goes to infinity. 

\smallskip
The precise definition is as follows:

\begin{definition}
\label{def1}
We say that a translate of~$\Lambda_B$ by~$B\bt$, where~$\bt\in\T_{L/B}$, is a \emph{good block}, or that the \emph{good block} event occurred in this translate if there exist two angles~$\theta^\star$ and~$\phi^\star$ such that:
\settowidth{\leftmargini}{(11)}
\begin{enumerate}
\item[(1)]
The angle~$\phi^\star$ satisfies either~$|\phi^\star|\le\kappa$ or~$|\phi^\star-180^\circ|\le\kappa$.
\item[(2)]
The collection of deviation angles~$\bvartheta=(\vartheta_{\br})$ defined from the angle variables~$\btheta=(\theta_{\br})$ and the angles~$\theta^\star$ and~$\phi^\star$ via \eqref{varthdef} obeys
\begin{equation}
\label{4.6}
|\vartheta_{\br}|<\Delta
\end{equation}
for all~$\br\in\T_L$.
\end{enumerate}
\end{definition}

Let~$\GG_0$ be the notation for good-block event with~$\phi^\star\approx0^\circ$ and let~$\GG_{180}$ be the good-block event for~$\phi^\star$ in the $\kappa$-neighborhood of~$180^\circ$. The complementary bad-block event will be denoted by~$\BB$. We remark that all these events depend only on the spin configuration (angle variables) in~$\Lambda_B$.

\begin{remark}
\label{rem2}
It is clear that if either~$\GG_0$ or~$\GG_{180}$ occurs (and if~$\kappa,\Delta\ll1$), then the spins in $\Lambda_B$ are indeed well-behaved in the sense of \twoeqref{2.2}{2.3} in Theorem~\ref{thm2.1}. Explicitly, if~$\br,\br'\in\Lambda_B$ is any pair of next-nearest neighbors, then~$\bS_{\br}\cdot\bS_{\br'}$ is very close to negative one. Moreover, on~$\GG_0$ we have $\bS_{\br}\cdot\bS_{\br'}\approx1$ when~$\br'=\br+\hate_x$ and $\bS_{\br}\cdot\bS_{\br'}\approx-1$ for~$\br'=\br+\hate_y$, while the opposite relations hold on~$\GG_{180}$. (Once~$\kappa,\Delta\ll1$, the requisite error is proportional to~$\kappa$ for next-nearest neighbors and to~$\Delta^2$ for the nearest neighbors.) Thus, the first step in obtaining \twoeqref{2.2}{2.3} will be to show that any particular block is of a given type of goodness with probability tending to one as~$\beta\to\infty$.
\end{remark}

Our goal is to use chessboard estimates to show that, with overwhelming probability, any given block is good and that, if one block is good with a known type of goodness, any other given block (regardless of the spatial separation) will exhibit the same type of goodness. As it turns out, on the basis of Theorem~\ref{thm4.1}, both of these will boil down to an efficient estimate of the quantity~$Z_{L,\beta}(\BB)$ defined in Section~\ref{sec4.1}. Unfortunately, here we will have to introduce a further partitioning: We let~$\BBE$ denote the event that, for some \emph{next-nearest} neighbor pair~$\br,\br'\in\Lambda_B$, we have
\begin{equation}
\label{4.7}
\bigl||\theta_{\br}-\theta_{\br'}|-\pi\bigr|\ge\frac\Delta{2B}.
\end{equation}
This event marks the presence of an energetic ``catastrophy'' somewhere in the block. As we will see, the complementary part of~$\BB$,
\begin{equation}
\label{4.8}
\BBSW=\BB\setminus\BBE
\end{equation}
denotes the situations where the energetics---and the spin-wave approximation---are good but where the configuration is not particularly near either of the spin-wave free-energy minima.

The event~$\BBSW$ will be further split according to the relative angle between the two near-Ne\'el states on even and odd sublattices. Specifically, we let~$\phi^\star_i$, $i=1,\dots,s$, be~$s$ angles uniformly spaced on the unit circle. Then we let~$\BBSW^{(i)}$ denote the event that the block~$\Lambda_B$ is bad but such that there exists an angle~$\theta^\star$ for which the deviation angles~$\bvartheta=(\vartheta_{\br})$ defined using~$\theta^\star$ and~$\phi^\star=\phi^\star_i$ satisfy $|\vartheta_{\br}|<\Delta$ at each~$\br\in\Lambda_B$. (Note that the second part is essentially the definition of the good block with the additional stipulation that~$\phi^\star=\phi^\star_i$ in part~(1) of Definition~\ref{def1}.) It remains to show that the~$\BBSW^{(i)}$ indeed cover~$\BBSW$:

\begin{lemma}
Let~$s$ be such that~$s\Delta>4\pi$. Then
\begin{equation}
\label{inkluze}
\BBSW\subseteq\bigcup_{i=1}^s\BBSW^{(i)}.
\end{equation}
\end{lemma}

\begin{proofsect}{Proof}
Consider a configuration of angle variables~$\btheta=(\theta_{\br})$ such that~$\BBSW$ occurs. Since this rules out the occurrence of~$\BBE$, we have
\begin{equation}
\pi-\frac\Delta{2B}<|\theta_{\br}-\theta_{\br'}|<\pi+\frac\Delta{2B}
\end{equation}
for any next-nearest neighbor pair $\br,\br'\in\Lambda_B$. But any two sites on the even sublattice in~$\Lambda_B$ can be reached in less than~$B$ steps and so $\theta_{\br'}$ for any even~$\br'\in\Lambda_B$ is within~$\Delta/2$ of~$\theta_0$ or~$\theta_0+\pi$, depending on the parity of~$\br'$ in the sublattice. Hence, the overall deviations from the appropriate Ne\'el state in direction~$\theta^\star=\theta_0$, where~$\theta_0$ is the variable at the torus ``origin,'' do not exceed~$\Delta/2$ throughout the even sublattice. Similar considerations apply to the odd sublattice where we use the positive $x$-neighbor of the origin to define the angle $\theta^\star+\phi^\star$.

It remains to show that the above implies that the spin configuration is contained in one of the events~$\BBSW^{(i)}$. Let~$i=1,\dots,s$ be the unique index such that~$\phi_i^\star\le\phi^\star<\phi_{i+1}^\star$, where $\phi_{s+1}^\star$ is to be interpreted as~$\phi_1^\star$. Then $|\phi^\star-\phi^\star_i|<2\pi/s$ which by our assumption is less than~$\Delta/2$. Consequently, all spins on the even sublattice are within~$\Delta$ of either~$\theta^\star$ or~$\theta^\star+\pi$, depending on the parity, while those on the odd sublattice are within~$\Delta$ of either~$\theta^\star+\phi_i^\star$ or~$\theta^\star+\phi_i^\star+\pi$, again depending on the parity. In particular, the event~$\BBSW^{(i)}$ occurs, thus proving \eqref{inkluze}.
\end{proofsect}

\subsection{Proofs of Theorems~\ref{thm2.1} and~\ref{thm2.2}}
\label{sec4.3}\noindent
As alluded to in the paragraph before \eqref{4.7}, the computational part of the proof boils down to estimates of the partition functions for events~$\BBE$ and~$\BBSW$. These will be provided in next two lemmas. We begin with the event~$\BBE$:

\begin{lemma}
\label{lemma4.4}
There exists~$\delta>0$ and constants~$c_2,c_3\in(0,\infty)$ such that if $\beta J\in(0,\infty)$ and $\Delta\in(0,1)$ satisfy the bounds \eqref{3.10}, then we have
\begin{equation}
\label{4.11}
\limsup_{L\to\infty}\Bigl(\frac{Z_{L,\beta}(\BBE)}{Z_{L,\beta}}\Bigr)^{(B/L)^2}\le4B^2
(c_3\beta J)^{B^2/2}e^{-c_2\beta J\Delta^2/B^2}.
\end{equation}
\end{lemma}

\begin{proofsect}{Proof}
When~$\BBE$ occurs, the exists a next-nearest neighbor bond in~$\Lambda_B$ where the associated angle variables satisfy \eqref{4.7}. An easy calculation shows that the energy this bond contributes to the Hamiltonian in \eqref{3.1}---note that the latter assigns zero energy to the Ne\'el ground states---exceeds the~$J$-multiple of
\begin{equation}
\label{4.12}
1+\cos\bigl(\pi-\tfrac\Delta{2B}\bigr)=2\sin^2\bigl(\tfrac\Delta{2B}\bigr).
\end{equation}
Bounding the sine from below by a linear function, which is justified because $\Delta/B\le\pi$, the right-hand side is not less than a numerical constant times~$(\Delta/B)^2$. We thus get 
\begin{equation}
\label{UBD}
Z_{L,\beta}(\BBE)^{(B/L)^2}\le4B^2
e^{-c_2\beta J\Delta^2/B^2},
\end{equation}
where~$c_2\in(0,\infty)$ is a constant and where~$4B^2\ge2B(B+1)$ bounds the number of ways to choose the ``excited'' bond in each translate of $\Lambda_B$.

Our next task is to derive a lower bound on the full partition function. A simple way to get such a bound is to insert the indicator that all angle variables~$\theta_\br$ are within~$\Delta$ of one of the spin-wave free energy minima, say,~$0^\circ$. This gives
\begin{equation}
Z_{L,\beta}\ge \Bigl(\frac{2\pi}{\beta J}\Bigr)^{L^2/2}e^{-L^2F_{L,\Delta}(0^\circ)},
\end{equation}
where~$F_{L,\Delta}$ is as in \eqref{3.7}.
Fix~$\epsilon>0$ and let $\delta>0$ be as in Theorem~\ref{thm3.2}. Then our assumptions on~$\beta$,~$\Delta$ and~$\delta$ and the conclusion \eqref{3.8} tell us that
\begin{equation}
\liminf_{L\to\infty}\,(Z_{L,\beta})^{1/L^2}\ge\Bigl(\frac{2\pi}{\beta J}\Bigr)^{1/2}
e^{-F(0^\circ)-\epsilon}.
\end{equation} 
Let us write the right-hand side as~$(c_3\beta J)^{-1/2}$, where~$c_3$ is a positive constant independent of~$\beta$ and~$\Delta$. Raising this bound to the~$B^2$ power and combining it with \eqref{UBD} the bound \eqref{4.11} is now proved. 
\end{proofsect}

Next we will attend to a similar estimate for the event~$\BBSW$:

\begin{lemma}
\label{lemma4.5}
For each~$\kappa\ll1$ and each~$\gamma\in(0,2)$ there exist numbers~$\rho(\kappa)>0$ and~$\delta>0$ such that if~$\Delta\ll\kappa$ and if~$\beta J$,~$\Delta$ and~$\delta$ satisfy the bounds in \eqref{3.10}, then
\begin{equation}
\label{4.13}
\limsup_{L\to\infty}\,\Bigl(\frac{Z_{L,\beta}(\BBSW)}{Z_{L,\beta}}\Bigr)^{(B/L)^2}\le 
8\pi\Delta^{-1} e^{-\rho(\kappa) B^2}.
\end{equation}
\end{lemma}

\begin{proofsect}{Proof}
Let~$\phi^\star_i$, $i=1,\dots,s$, be~$s$ angles uniformly spaced on the unit circle. Suppose that~$s$ and~$\Delta$ satisfy~$4\pi<s\Delta<8\pi$.
In light of the decomposition \eqref{inkluze} and the subadditivity property from Lemma~\ref{lemma4.2}, it suffices to show that, under the conditions of the lemma,
\begin{equation}
\label{4.15aa}
\limsup_{L\to\infty}\,\Bigl(\frac{Z_{L,\beta}(\BBSW^{(i)})}{Z_{L,\beta}}\Bigr)^{(B/L)^2}\le 
e^{-\rho(\kappa) B^2}
\end{equation}
for every~$i=1,\dots,s$.

First we note that for~$\phi^\star_i$ nearer than~$\kappa-\Delta$ to either~$0^\circ$ or~$180^\circ$ we automatically have~$\BBSW^{(i)}\subset\GG_0\cap\GG_{180}$. But then~$\BBSW^{(i)}=\emptyset$ because the event~$\BBSW^{(i)}$ is a subset of~$\BB$. By our assumption that~$\Delta\ll\kappa$ we just need to concentrate only on~$i=1,\dots,s$ such that~$\phi^\star_i$ is at least, say,~$\kappa/2$ from~$0^\circ$ or~$180^\circ$. Here we will use that~$Z_{L,\beta}(\BBSW^{(i)})$ is exactly the $(\frac{2\pi}{\beta J})^{L^2/2}$ multiple of the integral in \eqref{3.7} with~$\phi^\star=\phi^\star_i$, while~$Z_{L,\beta}$ can be bounded from below by a similar quantity for~$\phi^\star=0^\circ$,~i.e.,
\begin{equation}
\Bigl(\frac{Z_{L,\beta}(\BBSW^{(i)})}{Z_{L,\beta}}\Bigr)^{1/L^2}
\le\exp\bigl\{-F_{L,\Delta}(\phi^\star_i)+F_{L,\Delta}(0^\circ)\bigr\}.
\end{equation}
Let now~$\epsilon>0$---whose size is to be determined momentarily---and choose~$\delta>0$ so that Theorem~\ref{thm3.2} holds. Then the quantities $F_{L,\Delta}(\phi^\star)$ on the right-hand side are, asymptotically as~$L\to\infty$, to within~$\epsilon$ of the actual spin-wave free energy. Hence, we will have
\begin{equation}
\limsup_{L\to\infty}\,\Bigl(\frac{Z_{L,\beta}(\BBSW^{(i)})}{Z_{L,\beta}}\Bigr)^{1/L^2}
\le\exp\bigl\{-F(\phi^\star_i)+F(0^\circ)+2\epsilon\bigr\}.
\end{equation}
This proves \eqref{4.15aa} with~$\rho(\kappa)$ given as the minimum of~$F(\phi^\star_i)-F(0^\circ)-2\epsilon$ over all relevant~$i$. To show that~$\rho(\kappa)$ is positive for~$\kappa\ll1$, we first recall that Theorem~\ref{thm3.3} guarantees that~$F(\phi^\star)$ is minimized only by~$\phi^\star=0^\circ,180^\circ$. Since all of the relevant~$\phi_i^\star$ are bounded away from these minimizers by at least~$\kappa/2$, choosing $\epsilon=\epsilon(\kappa)>0$ sufficiently small implies $\rho(\kappa)>0$ as desired.
\end{proofsect}

Apart from the above estimates, we will need the following simple observation:

\begin{lemma}
\label{lemma4.6}
Let~$\kappa\ll1$. Then for any be two neighboring vectors~$\bt_1,\bt_2\in\T_{L/B}$,
\begin{equation}
\tau_{\bt_1}(\GG)\cap\tau_{\bt_2}(\GG)=
\bigl(\tau_{\bt_1}(\GG_0)\cap\tau_{\bt_2}(\GG_0)\bigr)
\cup
\bigl(\tau_{\bt_1}(\GG_{180})\cap\tau_{\bt_2}(\GG_{180})\bigr).
\end{equation}
In other words, any two neighboring good blocks are necessarily of the same type of goodness.
\end{lemma}

\begin{proofsect}{Proof}
Since~$\GG=\GG_0\cup\GG_{180}$, the set on the right is a subset of the set on the left. The opposite inclusion is a simple consequence of the fact that neighboring blocks share a line of sites along their boundary. Indeed, suppose the shared part of the boundary is parallel with the~$y$ axis. For~$\kappa\ll1$, Definition~\ref{def1} requires that the neighboring boundary spins are nearly aligned in a~$\GG_{180}$-block and nearly antialigned in a~$\GG_0$-block. Hence, the type of goodness must be the same for both blocks.
\end{proofsect}

Now we are ready to prove our main result:

\begin{proofsect}{Proof of Theorem~\ref{thm2.1}}
As is usual in the arguments based on chessboard estimates, the desired Gibbs states will be extracted from the torus measure~$\BbbP_{L,\beta}$ defined in Section~\ref{sec4.1}. Throughout the proof we will let~$\beta$ be sufficiently large and let~$\Delta$ scale as a (negative) power of~$\beta$ with exponent strictly between~$1/3$ and~$1/2$, and~$B$ grow slower than any power of~$\beta$,~e.g., as in
\begin{equation}
\label{DBass}
\Delta=\beta^{-\frac5{12}}\quad\text{and}\quad B=\log\beta.
\end{equation}
We note that these relations (eventually) ensure the validity of the bounds \eqref{3.10} for any given~$\delta>0$ and thus make the bounds in Lemmas~\ref{lemma4.4}-\ref{lemma4.5} readily available.

First we will show that in any typical configuration from~$\BbbP_{L,\beta}$ most blocks are good.
Let~$\eta_L$ denote the sum of the ratios on the left-hand side of \eqref{4.11} and \eqref{4.13}, i.e.,
\begin{equation}
\eta_L=\Bigl(\frac{Z_{L,\beta}(\BBE)}{Z_{L,\beta}}\Bigr)^{(B/L)^2}
+\Bigl(\frac{Z_{L,\beta}(\BBSW)}{Z_{L,\beta}}\Bigr)^{(B/L)^2},
\end{equation}
and let~$\eta=\limsup_{L\to\infty}\eta_L$.
By Theorem~\ref{thm4.1} and Lemma~\ref{lemma4.2}, the probability of a good block is then asymptotically in excess of~$1-\eta$.
On the basis of Lemmas~\ref{lemma4.4}-\ref{lemma4.5},~$\eta$ is bounded by the sum of the right-hand sides of \eqref{4.11} and \eqref{4.13} which under the assumptions from \eqref{DBass} can be made as small as desired by increasing~$\beta$ appropriately.

It remains to show that blocks with distinct types of goodness are not likely to occur in one configuration. To this end let us first observe that, once~$\kappa$ is small, no block can simultaneously satisfy both events~$\GG_0$ and~$\GG_{180}$. Invoking also Lemma~\ref{lemma4.6}, in any given connected component of good blocks the type of goodness is homogeneous throughout the component. (Here the notion of connectivity is defined via~$\T_{L/B}$, i.e., blocks sharing a line of sites in common, but other definitions would work as well.) We conclude that two blocks exhibiting distinct types of goodness must be separated by a closed surface (here $*$-connected) consisting of bad blocks. 

We will now employ a standard Peierls' estimate. For any~$\bt\in\T_{L/B}$ the event~$\GG_0\cap\tau_{\bt}(\GG_{180})$ is contained in the union of events that the respective blocks are separated by a~$*$-connected surface involving, say,~$m$ bad blocks. Using our choice of~$\eta$, Lemma~\ref{lemma4.2} and Theorem~\ref{thm4.1}, the probability of any surface of this size is bounded by~$\eta^m$. Estimating the number of such surfaces by~$c^m$, for some sufficiently large~$c<\infty$, and noting that~$m$ is at least~$4$, we get
\begin{equation}
\BbbP_{L,\beta}\bigl(\GG_0\cap\tau_{\bt}(\GG_{180})\bigr)\le\sum_{m\ge4}(c\eta)^m.
\end{equation}
Obviously, the right-hand side tends to zero as~$\eta\downarrow0$.

Thus, informally, not only are most blocks good, but most of them are of particular type of goodness. To finish the argument, we can condition on a block farthest from the origin to be, say, of~$\GG_{180}$-type. This tells us, uniformly in~$L$, that with overwhelming probability the block at the origin is of type~$\GG_{180}$ and similarly for the other type of goodness. The conditional state still satisfies the DLR condition for subsets not intersecting the block at the ``back'' of the torus. Taking the limit~$L\to\infty$ establishes the existence of two distinct infinite-volume Gibbs states which clearly satisfy \twoeqref{2.2}{2.3} with~$\epsilon(\beta)$ directly related to~$\eta$ and the various other parameters (cf Remark~\ref{rem2} in Section~\ref{sec4.2}).
\end{proofsect}

\begin{proofsect}{Proof of Theorem~\ref{thm2.2}}
This is, of course, just a Mermin-Wagner theorem. Indeed, the Hamiltonian \eqref{Ham} satisfies the hypotheses of, e.g., Theorem~1 in~\cite{ISV}, which prohibits breaking of any (compact) continuous internal symmetry of the model.
\end{proofsect}

\section*{Acknowledgments}
\noindent
The research of M.B. and L.C.~was supported by the NSF under the grant NSF~DMS-0306167; the research of S.K. was supported by the DOE grant DE-FG03-00ER45798.



\begin{thebibliography}{aaa}
\bibitem{Abanov}
Ar.~Abanov, V.~Kalatsky, V.L.~Pokrovsky and W.M.~Saslow, \textit{Phase diagram of ultrathin ferromagnetic films with perpendicular anisotropy}, Phys. Rev.~B \textbf{51} (1995) 1023--1038.

\bibitem{Alexander-Chayes}
K.S.~Alexander and L.~Chayes, \textit{Non-perturbative criteria for Gibbsian uniqueness}, Commun. Math. Phys. \textbf{189}  (1997),  no. 2, 447--464.

\bibitem{BCN1}
M.~Biskup, L.~Chayes and Z.~Nussinov, \textit{Orbital ordering in transition-metal compounds: I. The 120-degree model}, submitted.

\bibitem{BCN2}
M.~Biskup, L.~Chayes and Z.~Nussinov, \textit{Orbital ordering in transition-metal compounds: II. The orbital compass model}, in preparation.

\bibitem{BCG}
Ph.~Blanchard, L.~Chayes and D.~Gandolfo,
\textit{The random cluster representation for the infinite-spin Ising model: application to QCD pure gauge theory}, Nucl. Phys.~B \textbf{588} (2000) 229--252.

\bibitem{Bricmont-Slawny}
J.~Bricmont and J.~Slawny, \textit{Phase transitions in systems with a finite number of dominant ground  states}, J. Statist. Phys.~\textbf{54} (1989), no. 1-2, 89--161.

\bibitem{CKS}
L.~Chayes, R.~Koteck\'y and S.B.~Shlosman, \textit{Staggered phases in diluted systems with continuous spins}, Commun. Math. Phys. \textbf{189}  (1997),  no. 2, 631--640.

\bibitem{CSZ}
L.~Chayes, S.B.~Shlosman and V.A.~Zagrebnov, \textit{Discontinuity of the magnetization in diluted ${\rm  O}(n)$-models}, J.~Statist. Phys. \textbf{98} (2000), no. 3-4, 537--549.


\bibitem{Dimock-Hurd}
J.~Dimock and T.R.~Hurd, \textit{Sine-Gordon revisited}, Ann. Henri Poincar\'e \textbf{1} (2000) 499--541.

\bibitem{Dinaburg-Sinai}
E.I.~Dinaburg and Ya.G.~Sina\u\i, \textit{An analysis of ANNNI model by Peierls' contour method},  
Commun. Math. Phys.~\textbf{98} (1985),  no. 1, 119--144.

\bibitem{DS}
R.L.~Dobrushin and S.~Shlosman, \textit{Absence of breakdown of continuous symmetry in two-dimensional models of statistical physics}, Commun. Math. Phys. \textbf{42} (1975) 31--40.

\bibitem{DS2}
R.L.~Dobrushin and S.B.~Shlosman, \textit{Phases corresponding to minima of the local energy}, Selecta Math. Soviet.~\textbf{1}  (1981), no. 4, 317--338.

\bibitem{Dobrushin-Zahradnik}
R.L.~Dobrushin and M.~Zahradn\'{\i}k,
\textit{Phase diagrams for continuous-spin models: an extension of the  Pirogov-Sina\u{\i} theory},
In: R.L.~Dobrushin (ed.), \textit{Mathematical problems of statistical mechanics and dynamics},  pp.~1--123, Math. Appl. (Soviet Ser.), vol.~6,  Reidel, Dordrecht,  1986. 

\bibitem{DLS}
F.J.~Dyson, E.H.~Lieb and B.~Simon, 
\textit{Phase transitions in quantum spin systems with isotropic and nonisotropic interactions}, J.~Statist. Phys.~\textbf{18} (1978) 335--383.


\bibitem{vanEnter-Shlosman1}
A.C.D. van Enter and S.B. Shlosman, \textit{First-order transitions for $n$-vector models in two and more dimensions: Rigorous proof}, Phys. Rev. Lett. \textbf{89} (2002) 285702.

\bibitem{vanEnter-Shlosman2}
A.C.D. van Enter and S.B. Shlosman, \textit{Provable first-order transitions for nonlinear vector and gauge models with continuous symmetries}, mp-arc-04-121.

\bibitem{Fisher-Selke}
M.E.~Fisher and W.~Selke, \textit{Infinitely many commensurate phases in a simple Ising model}, Phys. Rev. Lett. \textbf{44} (1980) 1502--1505.

\bibitem{Fisher-Szpilka}
M.E.~Fisher and A.M.~Szpilka, \textit{Domain-wall interactions. II.~High-order phases in the axial next-nearest-neighbor Ising model}, Phys. Rev.~B \textbf{36} (1987) 5343--5362.

\bibitem{FILS1}
J.~Fr{\"o}hlich, R.~Israel, E.H.~Lieb and B.~Simon,
\textit{Phase transitions and reflection positivity. I.~General theory and long-range lattice models}, Commun. Math. Phys.~\textbf{62} (1978), no. 1, 1--34.

\bibitem{FILS2}
J.~Fr{\"o}hlich, R.~Israel, E.H.~Lieb and B.~Simon,
\textit{Phase transitions and reflection positivity. II.~Lattice systems with short-range and Coulomb interations}, J.~Statist. Phys.~\textbf{22} (1980), no. 3, 297--347.

\bibitem{FSS}
J.~Fr{\"o}hlich, B.~Simon and T.~Spencer, 
\textit{Infrared bounds, phase transitions and continuous symmetry breaking},
Commun. Math. Phys.~\textbf{50} (1976) 79--95.


\bibitem{Frohlich-Spencer}
J.~Fr{\"o}hlich and T.~Spencer, \textit{The Kosterlitz-Thouless transition in two-dimensional abelian spin  systems and the Coulomb gas}, Commun. Math. Phys. \textbf{81}  (1981),  no. 4, 527--602.

\bibitem{Georgii} 
H.-O.~Georgii, \textit{Gibbs Measures and
Phase Transitions}, de~Gruyter Studies in Mathematics, vol.~9,
Walter de Gruyter~\&~Co., Berlin, 1988.

\bibitem{Henley}
C.L.~Henley, \textit{Ordering due to disorder in a frustrated vector antiferromagnet}, Phys. Rev. Lett. \textbf{62} (1989) 2056--2059.

\bibitem{Holicky-Zahradnik}
P.~Holick\'y and M.~Zahradn\'{\i}k,
\textit{Stratified low temperature phases of stratified spin models: A general Pirogov-Sinai approach}, mp-arc 97-616. 

\bibitem{ISV}
D.~Ioffe, S.~Shlosman and Y.~Velenik, \textit{2D models of statistical physics with continuous symmetry: The case of  singular interactions},  Commun. Math. Phys. \textbf{226}  (2002),  no. 2, 433--454.

\bibitem{Kivelson-Fradkin-Emery}
S.A.~Kivelson, E.~Fradkin and V.J.~Emery, \textit{Electronic liquid-crystal phases of a doped Mott insulator}, Nature \textbf{393} (1998) 550--553.

\bibitem{Kosterlitz-Thouless}
J.M.~Kosterlitz and D.J.~Thouless, \textit{Ordering, metastability and phase transitions in two-dimensional systems}, J.~Phys.~C: Solid State Phys. \textbf{6} (1973) 1181-1203.

\bibitem{Kotecky-Shlosman}
R.~Koteck\'y and S.B.~Shlosman, \textit{First-order phase transitions in large entropy lattice models}, Commun. Math. Phys. \textbf{83}  (1982),  no. 4, 493--515.

\bibitem{Marchetti-Klein}
D.H.U.~Marchetti and A.~Klein, \textit{Power-law falloff in two-dimensional Coulomb gases at inverse temperatures~$\beta>4\pi$}, J.~Statist. Phys. \textbf{64} (1991) 135--162.

\bibitem{McBryan-Spencer}
O.A.~McBryan and T.~Spencer, \textit{On the decay of correlations in ${\rm SO}(n)$-symmetric  ferromagnets}, Commun. Math. Phys. \textbf{53}  (1977),  no. 3, 299--302

\bibitem{Mermin}
N.D.~Mermin, \textit{Absence of ordering in certain classical systems}, J.~Math. Phys. \textbf{8} (1967), no 5, 1061--1064. 

\bibitem{Mermin2}
N.D.~Mermin, \textit{The topological theory of defects in ordered media}, Rev. Mod. Phys. \textbf{51} (1979), no.~3, 591--648.

\bibitem{Mermin-Wagner}
N.D.~Mermin and H.~Wagner, \textit{Absence of ferromagnetism or antiferromagnetism in one- or two-dimensional isotropic Heisenberg models}, Phys. Rev. Lett. \textbf{17} (1966), no. 22, 1133--1136.

\bibitem{PSa}
S.A~Pirogov and Ya.G.~Sinai, \textit{Phase diagrams of classical lattice systems} (Russian), Theor. Math. Phys. \textbf{25} (1975), no. 3, 358--369.

\bibitem{PSb}
S.A~Pirogov and Ya.G.~Sinai, \textit{Phase diagrams of classical
lattice systems. Continuation} (Russian), Theor. Math. Phys.
\textbf{26} (1976), no. 1, 61--76.
  
\bibitem{Senya1}
S.~Shlosman, \textit{Phase transitions for two-dimensional models with isotropic short range interactions and continuous symmetry}, Commun. Math. Phys. \textbf{71} (1980) 207--212.

\bibitem{Senya2}
S.B.~Shlosman, \textit{The method of reflective positivity in the mathematical theory of phase  transitions of the first kind} (Russian), Uspekhi Mat. Nauk  \textbf{41} (1986), no. 3(249), 69--111.

\bibitem{Shender}
E.F.~Shender, \textit{Antiferromagnetic garnets with fluctuationally interacting sublattices},
Sov. Phys. JETP \textbf{56} (1982) 178--184 .

\bibitem{Villain}
J.~Villain, R.~Bidaux, J.-P.~Carton and R.~Conte, \textit{Order as an effect of disorder},
J. Phys. (Paris) \textbf{41} (1980), no.11, 1263--1272. 

\bibitem{Z1}
M.~Zahradn\'{\i}k, \textit{An alternate version of Pirogov-Sinai theory}, Commun. Math. Phys. \textbf{93} (1984) 559--581.

\bibitem{Zahradnik}
M.~Zahradn\'{\i}k, \textit{Contour methods and Pirogov-Sinai theory for continuous spin lattice  models}, In:~R.A.~Minlos, S.~Shlosman and Yu.M.~Suhov (eds.), \textit{On Dobrushin's way. From probability theory to statistical  physics},   pp.~197--220, Amer. Math. Soc. Transl. Ser.~2, vol.~198, Amer. Math. Soc., Providence, RI, 2000.

\end{thebibliography}
\end{document}